\documentclass[pdflatex,sn-mathphys-num]{sn-jnl}
\usepackage{graphicx}
\usepackage{amsmath}
\usepackage{url}
\begin{document}

\title[Sand sizes held by a pufferfish nest]{Sand sizes held by the ridge geometry of a pufferfish nest}

\author*[1,2]{\fnm{Yong Sung} \sur{Park}}\email{dryspark@snu.ac.kr}
\affil[1]{\orgdiv{Department of Civil, Urban and Environmental Engineering}, \orgname{Seoul National University}, \orgaddress{\city{Seoul}, \country{Republic of Korea}}}
\affil[2]{\orgdiv{Institute of Construction and Environmental Engineering}, \orgname{Seoul National University}, \orgaddress{\city{Seoul}, \country{Republic of Korea}}}

\abstract{Off southern Japan the white-spotted pufferfish builds a nest two meters across with 24--34 ridges radiating from a nearly flat center. Fine sand collects at the center while the male maintains the nest. From published three-dimensional scans of six nests, we compute how the ridges slow the near-bed water and which grain sizes they hold. The calculation does not use measurements of the nest sand. Its one free length is calibrated on reported results from a laboratory model and a flow simulation. Under ninetieth-percentile waves admitted to the site, the flow over the ridges would hold sand of 76--187 micrometers in their troughs. This range lies within the 50--300 micrometers reported at the center. The calculation assumes the ridges keep their built shape and the sand is not sorted again between the troughs and the center. Waves and tides move fine sand at the site for 5--13\% of the breeding season. In all six nests the ridges are steepest halfway across the ridge zone, where the range held is widest.}

\maketitle

\section{Introduction}
Animals building with sediment change the flow around them. A recent census for zoogeomorphology, the study of landforms made and modified by animals, counts more than 600 taxa with reported geomorphic effects in terrestrial and freshwater environments \cite{harvey2025}. It leaves marine and coastal environments outside its scope and finds the tropics and subtropics understudied \cite{harvey2025}. Marine builders whose effect on the flow has been measured include beds of many elements, such as mussel beds and oyster reefs \cite{lee2025,hitzegrad2025}. The white-spotted pufferfish \emph{Torquigener albomaculosus} builds its nest on the seabed off Amami-Oshima, Japan, at 28$^\circ$N. The nest is a single structure with a regular radial architecture, built by one male. The census would place the nest under its nest type, a landform type counted separately from the redd \cite{harvey2025}.

Males excavate a circular structure about 2~m across, with 24--34 radial ridges around a nearly flat central bed \cite{kawase2013,kawase2015,kawase2017,matsuura2015,matsuura2022}. In the surveys of June and July, construction took seven to nine days and ended with spawning \cite{kawase2013}. At completion the male draws the fine sand on the central bed into an irregular pattern with his anal fin. Fine sand accumulates on the central bed while the structure is built and maintained. After spawning the male tends the eggs on the central bed for five to six days, stirring the sand over them \cite{kawase2013,kawase2015}. He no longer maintains the ridges, which flatten \cite{kawase2013}. The fine sand at the center gives way to coarser sand soon after spawning \cite{kawase2016kaken}.

Kawase and colleagues, who identified the builder of the nests, proposed that the structure serves female mate choice and also gathers the fine sand the authors considered important to the choice \cite{kawase2013}. In their account the male stirs up the sand as he digs along each valley, the finer grains settle on the valley floors, and the ridges on either side keep the stirred grains from scattering. Water enters the upstream valleys toward the center whatever the current direction, and the male's outward passes carry the fine sand gathered in the valleys to the central bed \cite{kawase2013}. Kawase and colleagues also proposed three conditions for a nest site: sand of a grain size suitable for building, currents which transport fine sand, and water calm enough not to destroy the structure \cite[p.~40]{kawase2015}. Two- and three-dimensional flow solutions report velocity and bed shear stress lower over the central bed than over the ridges \cite{shameem2021,zhao2023,dai2024}. These solutions treat the flow as steady, one of them as laminar, use boundary-layer results at most in the wall treatment, and stop at the bed shear stress. The three conditions and the retention of fine sand have therefore not yet been given numbers. The forcing at the site has not entered the argument.

We measure the radial profiles of ridge height and wavelength on the three-dimensional scans of six completed nests made and deposited for open use by Kawase, Kitajima and Iwai \cite{kawase2022,kawase2022data}. The roughness the ridges present to the flow follows from their measured height. We bracket it between the roughness of ripples \cite{nielsen1992,grant1982} and an element-roughness rule obtained for oyster reefs \cite{lee2025}. We apply a wave--current bottom boundary-layer model over the measured profiles \cite{grant1979,glenn1987,goud1987}. The model gives the grain sizes staying in suspension over the ridge crests and settling in the troughs (Fig.~\ref{fig:object}b). At each radius these sizes form a range bounded at both ends. We compare the range with the sand sizes reported at the nest center, locate the radius at which it is widest, and compute the fraction of the breeding season in which waves and tides move fine sand at the site.

\section{Results}
\subsection{The landform measured}
Six completed nests, scanned on one day at one site and deposited as three-dimensional models \cite{kawase2022}, give the geometry (Fig.~\ref{fig:object} and Table~\ref{tab:morph}). Each has 27--30 ridges. Over the three radial intervals inside the ridge zone the dominant azimuthal order equals the number of ridges in five nests. In K11 it rises from 27 to 30 in the outermost of the three (Supplementary Table~1 and Methods). The number of ridges is therefore nearly constant with radius. The azimuthal wavelength is fixed by geometry, $\lambda(r) = 2\pi r/N$, and shrinks inward. The ridge height $\eta(r)$ is the one free profile (Fig.~\ref{fig:object}e). The ridge zone runs from $r_c$ = 434~mm, where the pattern begins, to $r_o$ = 953~mm, where the ridge amplitude falls below a fifth of its maximum. At the annular crest, $r$ = 0.85~m, the ridges are 2.3~cm high and 18.9~cm apart. Averaged over the ridge zone they are 1.9~cm high and 15.5~cm apart.

\subsection{Maintenance of the built geometry}\label{sec:maintenance}
Construction has been filmed from its early stage to completion \cite{kawase2013,kawase2017}, but the geometry on the scans could since have been adjusted by the flow. The ridge height and wavelength are similar to those of natural ripples \cite{lyne1990}. At the ninetieth percentile of the near-bed waves admitted to the site (Methods), the mobility number lies in the ripple range \cite{nielsen1992}. Measurements on the same scans show no such adjustment (Supplementary Note~1). The seabed around each nest has no ripples above the detection limit at wavelengths of 15--25~cm. The ridges run straight outward, with a phase drift of 9~mm against the 81~mm needed to move a ridge into its neighbor's place. Their crests follow the local radius to a median deviation of 5.2$^\circ$ (Supplementary Figs.~1 and 3). Without maintenance the ridges flatten within six days \cite{kawase2013}. A bedform formed by the forcing would persist under the same forcing. Leveling the ridges in six days requires a bedload transport barely above threshold, so the flattening does not constrain the magnitude of the forcing (Methods).

In the account of Kawase and colleagues the fish also sorts the sand \cite{kawase2013}. In the experiment reported by Kawase, a screw run at the speed at which the fish throws sand toward the center deposited sand of the field size from the site. The screw hardly moved a coarser sand from an entirely different site \cite{kawase2016kaken}. The two reported determinations made without fish \cite{kawase2013,shameem2021} concern the flow only and show that the geometry slows the near-bed flow at the center. The flow and the sand thrown by the fish can act together while the nest is maintained. The following calculation does not separate them.

\subsection{An intermittent forcing through a narrow aperture}\label{sec:forcing}
The site lies inside an embayment at the eastern entrance of the Oshima Strait. The directions open to the Pacific total 21.8$^\circ$, of which a contiguous 20.5$^\circ$ passes through a single 1.15~km gap (Supplementary Fig.~2a,b). Across the four depth intervals and the grain sizes considered, waves admitted through the gap exceed the threshold for fine sand for 2.0--5.7\% of the breeding season, April to September. The tidal current alone exceeds it for 2.7--6.9\% (Supplementary Fig.~2c). With the two combined in the wave--current boundary layer, the threshold is exceeded for 5--13\% of the season, 1.8--3.5 times the value for waves alone. The bracket of the tidal attenuation ratio, 0.22--0.27, moves the combined values by a factor of 0.68--1.7 across the depth intervals and grain sizes, to 3.4--19\%.

The one current reported from the observation area, about 0.8~m~s$^{-1}$ \cite{kawase2013}, is 1.4 times the largest reconstructed tide at the site. A tide scaled to reach 0.8~m~s$^{-1}$ raises the combined exceedance to at least 25--31\% for 0.15~mm sand (Supplementary Fig.~2c) and 22--34\% across grain sizes (Supplementary Note~3). In the fortnight before the six scanned nests were completed, no wave in either record crosses the threshold on its own (Supplementary Note~3). For waves alone, in three of the four depth intervals the median interval between episodes above threshold exceeds the six days in which an unmaintained structure flattens.

\subsection{Transmission to the center}\label{sec:transmission}
The transmission is the fraction of the near-bed speed outside the nest reaching the center. It depends on how rough the ridges are to the flow. Because the number of ridges is nearly constant with radius, the roughness of the ripple family (Methods) is $k_b = C N\eta^2/(2\pi r)$. The bed becomes rougher inward even where the ridge height is constant. In the half-scale model of Kawase and colleagues the current passes through the structure. It enters the center along the upstream valleys, leaves along the downstream valleys, and crosses the ridge tops and the central zone in its own direction. Its speed in the central zone is 24.2\% below the speed outside \cite{kawase2013}. We treat the ridge zone in one dimension along the path of this through-flow. $U(r)$ is the near-bed speed at distance $r$ from the center on the upstream side. The ridges present an azimuthally averaged effective roughness, because the orientation of the path relative to the ridges changes with the direction of the ambient flow. The speed decays because the bed stress removes momentum from a near-bed layer of thickness $\delta_w$ (Methods). Integrating the wake momentum deficit inward gives
\begin{equation}
U(r) = U_o \exp\!\left[-\frac{1}{2\delta_w}\int_r^{r_o} f_w \, \mathrm{d}r'\right],
\label{eq:transmission}
\end{equation}
with the wake thickness $\delta_w$ the only geometric coefficient. It scales with the effective obstacle height $h$, the elevation of the annular crest above the central bed, 71~mm (range 64--79) on the six scans (Methods).

In the steady-flow limit the transmission $T_u$ at the center has been determined twice without fish. The half-scale physical model gave 0.758 \cite{kawase2013}, at an unstated height. A two-dimensional steady computation gave 0.698 \cite{shameem2021}. A second two-dimensional computation \cite{dai2024} gives 0.125--0.25, which would need a wake thinner than the ridges, and is not used (Methods). With a friction factor inside the roughness bracket, calibration on the two values bounds the wake thickness to 36--82~mm, 0.5--1.2 times $h$ (Supplementary Fig.~4). The bracket spans both values at one wake thickness only in this interval. The width of the calibrated interval follows from the width of the roughness bracket, so the agreement calibrates the closure and does not test it independently.

\subsection{The confinement window}\label{sec:window}
Over each ridge the near-bed flow is faster across the crest than across the trough. A grain fine enough to stay in suspension over the crest but coarse enough to settle in the trough is held in the trough. The range of sizes so held is the confinement window (Fig.~\ref{fig:object}b). The slowing of the flow toward the center shifts both edges of the window together. The shielding factor at the inner edge is the friction velocity there relative to its value at the outer edge of the ridge zone. At $\delta_w = h$ it is 0.81--0.85 at the ninetieth-percentile forcing, and 0.67--0.87 across the calibrated interval of 36--82~mm. The contrast which controls grain sorting is found locally, between each ridge crest and its trough. To first order in potential flow over a wave-shaped bed (Fig.~\ref{fig:object}b; Methods),
\begin{equation}
\frac{u_{*,\mathrm{crest}}}{u_{*,\mathrm{trough}}}
= \frac{1 + \pi\eta/\lambda}{1 - \pi\eta/\lambda},
\label{eq:contrast}
\end{equation}
which we use as a geometric index of the contrast. Over ridges as steep as these the flow separates and the equation does not predict the stress. Any estimate of the contrast increasing with the local steepness places the widest window at the radius of greatest steepness. The cut-off grain sizes separating bedload from suspension (Rouse criterion, Methods), evaluated at crest and trough, define the confinement window $D^*_\mathrm{trough} < D < D^*_\mathrm{crest}$. Its width is $\sqrt{(1+\pi\eta/\lambda)/(1-\pi\eta/\lambda)}$. On the six-nest mean profile this width is at most 1.73 in the Stokes range of settling and 1.79--1.86 with the full settling law at the ninetieth percentile.

On the mean profile the steepness $\eta/\lambda(r)$, a property of the built geometry, peaks at 0.159 at $r$ = 668~mm. It falls to 0.063 and 0.035 at the inner and outer edges (Fig.~\ref{fig:radial}a,b). In each of the six nests the steepness peaks between 618 and 733~mm (683 $\pm$ 46~mm, SD), at 0.45--0.53 of the distance from $r_c$ to $r_o$ (Fig.~\ref{fig:radial}c). Both edges of the ridge zone are amplitude crossings at 20\% of the maximum, so an interior maximum follows from the definition. Six independent random profiles cut the same way fall as close together as the six nests with a probability of about $5\times10^{-4}$ to 0.08, depending on the smoothness assumed (Supplementary Fig.~5 and Supplementary Note~1). The agreement rests partly on K12, whose second maximum at 538~mm reaches 0.99 of the peak. The ridge height itself peaks at 0.46--0.81 of the zone.

A ripple field of constant amplitude and a self-similar grader (Methods) both have $\lambda \propto r$, so both are steepest at the inner edge. Their windows decrease outward, from 1.99 to 1.32 and from 2.58 to 1.27 (Fig.~\ref{fig:radial}b). An unbounded equilibrium wave-ripple field at the ninetieth-percentile forcing would have a steepness of 0.062--0.125 by the empirical relation of Nielsen \cite{nielsen1992} (scatter about $\pm$30\%), uniform in radius. Its window would be 1.22--1.52 at every radius. The measured window lies inside this range over 27\% of the ridge zone and above it over 66\%. Replacing the ratio in equation~(\ref{eq:contrast}) by its square root, which halves the contrast in logarithmic terms, narrows the widest window to 1.32 at the same radius.

We evaluate the window only at the ninetieth-percentile forcing of the four depth intervals, a state reached for 1.1\% of the season (Methods). At this forcing the crest friction velocity exceeds the threshold of motion for every grain in the window at every radius (Methods). The predicted range is 76--187~$\mu$m, centered in very fine sand (Fig.~\ref{fig:window}). Grain size enters only through the grain roughness of the bed (Methods). The 21--25~m interval gives 82--149~$\mu$m. The calibrated wake thickness of 36--82~mm gives 72--188~$\mu$m. With the ratio in equation~(\ref{eq:contrast}) replaced by its square root the range is 93--169~$\mu$m. At the radius of the widest window the trough stays below the threshold of motion in the 21--25 and 23--28~m intervals. In the 18--23 and 13--16~m intervals it exceeds the threshold by 2 and 21\%, and grains of the window move along the trough without being suspended. In all four the lower edge is a settling limit. At median forcing the crest stays below the threshold for every grain of the window in the 23--28~m interval (Fig.~\ref{fig:window}b). It exceeds the threshold for the finest grains of the window at about half the radii in the 21--25~m interval, and for the coarsest at most radii in the 18--23 and 13--16~m intervals. The window computed at median forcing, 44--110~$\mu$m across the four intervals, is not used. On a flat bed under the same ninetieth-percentile forcing the suspension cut-off is a single size, 117--154~$\mu$m across the four depth intervals. The window spans 0.65--1.21 times this cut-off. The step from the troughs to the central bed, which Kawase and colleagues attribute to flow into the upstream valleys and to the male's outward passes \cite{kawase2013}, is not computed. The comparison with the reported sizes assumes this step does not sort the sand again.

A laser particle-size analysis of sand from the parts of the nest is reported by Kawase in outline in a public grant report \cite{kawase2016kaken}. It found sand of 50--300~$\mu$m accumulating gradually at the center and giving way to coarser sand soon after spawning. It also found the sand similar in size at sites whose beds differ in grain-size composition. The predicted range lies within the reported range, whose upper end is about 1.6 times as large. The report gives neither a distribution nor a radius within the central bed. The flat-bed cut-off also falls within the reported range. The screw experiment comes from the same report \cite{kawase2016kaken}, so the reported sizes are expected both from the window and from the speed at which the fish throws sand. The loss of the fine sand soon after spawning \cite{kawase2016kaken}, when the male stops maintaining the ridges and begins to stir the sand over the eggs \cite{kawase2013,kawase2015}, does not separate the two either. The window predicts a held range in the troughs, varying with radius and present only when the bed is mobile. The speed at which the fish throws sand sets a size for the sand delivered to the center and makes no prediction for the troughs.

\section{Discussion}
The window gives a size to the retention of fine sand proposed by Kawase and colleagues \cite{kawase2013,kawase2022}. It would act together with their mechanism of stirring and outward passes. If the geometry sorts the sand, two of the three conditions they proposed for a nest site \cite[p.~40]{kawase2015} receive estimated values. The condition on grain size becomes a requirement on the supply. At this site the bed must contain sand of 76--187~$\mu$m. The condition on transporting currents becomes the fraction of the season in which the forcing moves fine sand, 5--13\%. The third, calmness, remains their observation that an unmaintained structure flattens within six days. Any radial structure built with a fixed number of elements becomes steeper inward wherever its element height is constant. The closure places the widest window where the steepness of the structure, $N\eta(r)/2\pi r$, is greatest.

Female salmon digging a redd winnow fine sediment from the gravel \cite{kondolf2000,tonina2009}. In a three-dimensional model of a redd the winnowed gravel raises the intragravel flow and the dissolved oxygen around the buried eggs \cite{tonina2009}. In the pufferfish nest fine sand collects at the center, where the eggs are deposited on the sand \cite{kawase2013,kawase2015}. Whether the fine sand serves the eggs has not been reported, and the present calculation bears only on its size. The two flow determinations did not include sand, and no observation yet shows the structure holding fine sand without the male. Building moves at least 54--96~kg of grains per nest. At the two sites, each surveyed over about $10^4$~m$^2$, at least 4--18 nests were observed in one season \cite{kawase2015}. One season therefore reworks at least 0.1--0.5\% of the seabed.

Similar nests, with 22--27 ridges, have been found at a second site on the southern North West Shelf of Australia, in 129--137~m of water \cite{bond2020}. Individuals of a pufferfish of the genus \emph{Torquigener} were seen near several of the 22 nests \cite{bond2020}. Video of a further nest nearby suggests the same material arrangement: fine sediment and a smooth surface at the center, with coarse, light-colored sediment exposed on the ridges and the outer ring \cite{bond2020}. The observation is a visual reading of video, and no size distribution was measured. The Japanese nests lie close to the coast in 13--28~m of water \cite{kawase2013,kawase2022,matsuura2022}, with single nests at 30 and 32~m \cite{matsuura2022,sonoyama2022}. Where the setting is reported they lie in bays, mostly on sand and in one case on a sand--mud bottom \cite{kawase2015,matsuura2022,sonoyama2022}. The Australian site is on the open shelf about 130~km from the coast \cite{bond2020}. We therefore do not treat any forcing or setting as a requirement. The Australian substrate is carbonate-dominated \cite{ausseabed}, so the assumed quartz density and sphericity do not apply there.

The geometry comes from six nests completed together on one day of one spawning cycle at depths of 13--16 and 21--25~m off one embayment \cite{kawase2022}. Earlier surveys off Seisui and Katetsu report 24--32 ridges for ten nests and outer radii of 79--105~cm for six \cite{kawase2013}. They do not give a ridge-height profile. The radial steepness profile has therefore not been examined at other sites or in other seasons. Propagation inside the embayment is not computed. The tidal attenuation ratio is read from a published figure. The one current reported from the observation area requires a ratio of at least 0.35, above the bracket of 0.22--0.27. The window is evaluated at one forcing state, the ninetieth percentile of the waves admitted through the aperture (Methods). The window under the combined wave--current stress is not computed. The Rouse criterion is evaluated with a settling velocity held constant, which turbulence can modify (Methods). A modification uniform in the settling parameter shifts the predicted range and does not move the radius of the widest window. A modification differing between ridge crest and trough by a factor $f$ would close the window at the zone edges for $f$ of about 1.2--1.5 and at the widest radius for $f$ of about 3 (Supplementary Note~2). The Rouse-number convention moves the absolute range by a factor of 0.83--1.55 and the radius of the widest window by at most 5~mm. Over the steepest ridges suspension is convective. The convection--diffusion model of Nielsen \cite{nielsen1992} gives 69--159~$\mu$m but resolves neither crest nor trough. The width of the window therefore rests on the diffusive criterion (Supplementary Note~2). We found no published position-resolved size distribution from within a nest. The grain size of the bed surrounding the nests at the site has not been reported. The model is one-dimensional in radius. Azimuthal channeling along the troughs, at least a two-dimensional problem, would need a sector-resolved three-dimensional solution.

The two determinations do not distinguish the two roughness families, because they fix the ratio $f_w/\delta_w$. The vertical structure of the velocity profile over one ridge separates the families. Either a ridge-resolving three-dimensional solution or oscillatory-flow particle image velocimetry would supply it. The sign of the steady streaming is predicted for progressive waves over long-crested ripples \cite{holmedal2009} and not for an axisymmetric radial array under combined forcing. The observed flow enters the center along the upstream valleys \cite{kawase2013}. Transmission and grading are derived for a steady forcing. The forcing at the site is intermittent and is weakest during the stratified summer, when stratification lowers the tidal energy in the strait by about a third \cite{hamasoe2019}. An accumulation model driven by the measured forcing record would show how grading accumulates under such a forcing.

\section{Conclusions}
The measured ridge geometry of a pufferfish nest is sufficient to hold sand of the sizes reported at its center under ninetieth-percentile waves at the site. The range held, 76--187~$\mu$m, lies within the reported 50--300~$\mu$m. Its width follows the ridge steepness, which peaks at mid-radius in all six nests. The result holds while the ridges keep their built shape and if the sand is not sorted again on its way from the troughs to the center. Verification against measured size distributions from known positions within a nest is left for future research.

\section{Methods}
\subsection{Morphometry}
Six completed nests, scanned by structure from motion off Katetsu, Amami-Oshima
($28^\circ07'53''$N, $129^\circ20'36''$E) on 15 June 2016, were retrieved from the deposited figshare collection under a CC0 license
\cite{kawase2022data}. The six reduced models have about 2.5 million vertices each at a mean
spacing near 2.7~mm. Each model was scaled by its authors from a single 30-cm ruler set outside the nest \cite{kawase2022}. Each cloud was
detrended, centered on the axis of the radial pattern, and resampled onto polar grids at 5~mm in radius.

The number of ridges was taken as the order of the dominant azimuthal harmonic. It is
27--30 over the six nests with a mean of 28.2 (Table~\ref{tab:morph}).
Ridge height was measured as twice the amplitude of the $N$-th azimuthal harmonic.

Three radii are kept apart by definition: $r_c$, the inner end of the pattern; $r_\mathrm{ring}$, the annular crest; and $r_o$, the outer end, where the ridge amplitude falls below 20\% of its maximum. The surface relief is the one quantity on which the extraction can be checked against the published description of the same nests. The extracted 112~mm (range 100--126) overlaps the published 106~mm (range 97--113).

The effective obstacle height $h$ used in the transmission model is the elevation of the azimuthally averaged annular crest above the mean elevation of the central bed inside $r$ = 0.15~m, measured on each scan. It is 70.9~mm (64.2--79.3, $n$ = 6; Table~\ref{tab:morph}). The crest stands 41~mm (34--53) above the bed outside the structure and the center lies 30~mm (11--40) below it.

The bed outside the ridge zone was tested for bedforms on the same scans (Supplementary Note~1).

The volume of sediment moved in building a nest was obtained by integrating the gridded elevation inside $r_o$ relative to the ambient bed at $r \geq$ 1.35~m, separately for the parts below (excavated) and above (piled) the ambient level. The volume moved is taken as the larger of the two, because sand dug from the troughs is piled on the crests. It is a lower bound because the scans record only the net relief. Building a nest moves at least 34--60~L (mean 45~L, $n$ = 6), about 4.5~cm$^3$ per digging event, the volume moved divided by the roughly 10,000 excavations of a build \cite{mizuuchi2018}. Leveling a sinusoidal ridge of height $\eta$ and wavelength $\lambda$ requires a volume per unit ridge length $\eta\lambda/4\pi = 3.4\times10^{-4}$~m$^2$ to cross the zero-crossing lines. Delivering this volume in six days requires a bedload transport barely above threshold. The mass and the seabed fraction in the Discussion take a porosity of 0.4, a grain density of 2650~kg~m$^{-3}$ and the area within the six-nest mean $r_o$.

\begin{table}[t]
\centering
\caption{Per-nest morphometry, extracted from the deposited scans (nest identifiers
follow the deposited file names). $\Delta z_\mathrm{azi}$ is the relief of the azimuthally
averaged profile between the annular crest and the inner moat; $\Delta z_\mathrm{surf}$ is the surface relief between the 0.5th
and 99.5th elevation percentiles; $a_N$ is the peak amplitude of the $N$-th
harmonic, so the maximum ridge height is about $2a_N$; $h$ is the elevation of the annular crest above the central bed.}
\small
\begin{tabular}{lcccccccc}
\hline
Nest & $N$ & $r_c$ & $r_\mathrm{ring}$ & $r_o$ &
$\Delta z_\mathrm{azi}$ & $\Delta z_\mathrm{surf}$ & $a_N$ & $h$ \\
& & [mm] & [mm] & [mm] & [mm] & [mm] & [mm] & [mm] \\
\hline
K1 & 27 & 473 & 883 & 968 & 86.7 & 125.8 & 11.6 & 74.3 \\
K2 & 28 & 438 & 813 & 903 & 84.2 & 109.2 & 21.1 & 64.2 \\
K11 & 27 & 383 & 813 & 898 & 89.0 & 113.3 & 13.2 & 79.3 \\
K12 & 29 & 448 & 893 & 1013 & 81.6 & 106.1 & 11.2 & 74.9 \\
K13 & 28 & 443 & 833 & 948 & 73.8 & 99.5 & 13.7 & 68.6 \\
K14 & 30 & 423 & 873 & 988 & 84.3 & 118.8 & 18.0 & 64.3 \\
\hline
Mean & 28.2 & 434 & 851 & 953 & 83.3 & 112.1 & 14.8 & 70.9 \\
\hline
\end{tabular}
\label{tab:morph}
\end{table}

\subsection{Wave--current bottom boundary layer}
Near-bed flow was closed with the Grant--Madsen--Glenn wave--current boundary layer
\cite{grant1979,glenn1987,goud1987}. The combined-flow factor $f_{cw}$ of the explicit Madsen form is used for the boundary layer. The angle between the wave direction transmitted through the aperture and the
axis of the reversing tidal current is 33$^\circ$.

Skin friction and total roughness are kept apart throughout. Thresholds of motion and
bedload use the grain roughness $2.5 D_{50}$ \cite{lee2024esp}. Transmission uses the total roughness. The two differ by about a factor
of six at this site.

The threshold is the critical Shields parameter of Soulsby and Whitehouse~\cite{soulsby1997threshold},
$\theta_{cr} = 0.30/(1 + 1.2 D_*) + 0.055\,[1 - \exp(-0.020 D_*)]$, with $D_*$ the
dimensionless grain diameter. The exceedance fractions quoted above are computed from this threshold through the combined-flow stress. Where a wave-only threshold velocity is needed,
we use the closed form of Komar and Miller~\cite{komar1973} for grains finer than 0.5~mm. The
coefficient is 0.30, recommended by those authors as the mean of their two calibrations.

\subsection{Roughness closure}
The ridges are treated as a macro-roughness, with two closures used in parallel as a bracket, neither of them fitted. The first is the 7$\Delta_o$ rule of Lee et al.~\cite{lee2025}, $k_s = 7\Delta_o$, with $\Delta_o$ the height of the roughness element. We write $k_b$ for the bed roughness of either closure. The rule was obtained for oyster-reef roughness from velocity profiles measured in steady and oscillatory flow. The coefficient is 6.6 (95\% confidence interval 5.9 to 7.3) and 6.7 (4.3 to 9.0). The two intervals overlap, so one coefficient serves both flows, as combined wave--current forcing requires. We assume $\Delta_o$ equals the crest-to-trough relief of the ridges. The rule was obtained on three-dimensional reef elements and is transposed to long ridges, which is one reason the second closure is kept. The second closure, included because the ridges are also bedforms, is the ripple family. Its bed roughness is $C\eta(\eta/\lambda)$ with $C$ between 7.4 and 27.7 across six published values \cite{nielsen1992,grant1982}. Across the bracket the roughness length changes by a factor of 7.8. The roughness bracket is converted to a wave friction factor in the explicit Madsen form for the transmission and the window (0.043--0.114 across the bracket at an orbital excursion of 0.32~m, the reference state $u_b$ = 0.20~m~s$^{-1}$ and $T$ = 10~s at which the bracket is tabulated, a factor of 2.7). In the steady calibration the same friction factor serves as the bulk friction factor of the current over the ridges. Because the two determinations fix only $f_w/\delta_w$, the calibrated interval of 36--82~mm is tied to this reference state. Transmission is reported over the whole bracket. The window is evaluated with $C$ = 27.7, the value of Grant and Madsen \cite{grant1982}. Across the bracket the window moves by at most 4\%.

At the rough end of the bracket and the median forcing the relative roughness $A/k_b$, with $A$ the orbital excursion, falls to 0.73, outside the range over which the friction-factor fits were established. The determination we transpose was itself made for $A/k_b$ below 10 \cite{lee2025}. We use the closure only as a bulk friction factor. In the steady limit the two determinations fix the ratio $f_w/\delta_w$, so an error in $f_w$ at the rough end is absorbed in the wake thickness. The window is set by skin friction on the grain roughness $2.5D_{50}$. The total roughness enters only through the shielding factor. The radius of the widest window does not move across the bracket by more than the 5-mm grid spacing. An independent measurement of the
wave friction factor of biogenic roughness admits the whole bracket
\cite{hitzegrad2025}, so it does not discriminate between the two families.

\subsection{Transmission}\label{sec:methods-transmission}
The inward decay of near-bed flow was obtained by integrating the wake momentum deficit
over the measured ridge zone. The bed stress $\tfrac12\rho f_w U^2$ decelerating a layer of thickness $\delta_w$, $\rho\delta_w U\,\mathrm{d}U/\mathrm{d}r = \tfrac12\rho f_w U^2$, gives $\mathrm{d}U/\mathrm{d}r = f_w U/2\delta_w$, whose integral is equation~(\ref{eq:transmission}). The flow is not a radial sink, so the balance omits a continuity term in $2\pi r$. The thickness is the depth over which the ridge array spreads its
momentum deficit and is distinct from the oscillatory boundary layer. Within each closure, the wake thickness at which the two determinations are met exceeds the wave boundary-layer length scale $\kappa u_*/\omega$ for the same conditions by a factor of 1.3--2.4. An internal boundary layer, in which the ridges only raise the roughness of a growing logarithmic layer, gives an attenuation of 2\%, an order of magnitude below both determinations. The attenuation in both determinations therefore requires the form drag of the ridges to act over a layer thicker than the boundary layer, as in the wake balance. In the steady limit the
expression is calibrated on the two determinations (Results). The steady value of the two-dimensional computation is taken as the square root of the ratio of the mean bed stresses the authors report on the central bed and on the surrounding bed. The square root equals the ratio of near-bed speeds where the friction coefficient is the same on both. The computation used a cross-section of a model reconstructed from video at an estimated scale \cite{shameem2021,kawase2022}. The half-scale model has a diameter of 1~m, or 2~m at full scale, against 1.80--2.03~m for the scanned nests. Its ridge count and relief are not given \cite{kawase2013}. The physical model is three-dimensional and represents the whole structure. The current speed used and the height of the measurement are not given \cite{kawase2013}. The second two-dimensional computation \cite{dai2024}, 0.125--0.25, would need a wake 5--21~mm thick, thinner than the 23~mm height of the ridges at the annular crest.

\subsection{Grading and the confinement window}\label{sec:methods-grading}
Suspension was separated with a Rouse criterion. With an eddy viscosity $\kappa u_* z$ and a sediment diffusivity equal to the eddy viscosity divided by the turbulent Schmidt number 0.74 of the combined boundary layer \cite{goud1987}, the suspended concentration decays with height as $z^{-0.74\,w_s/(\kappa u_*)}$. We place the boundary between bedload and suspension at an exponent of 1, so the cut-off grain size satisfies $w_s(D^*) = \kappa u_*/0.74$, a Rouse number $w_s/(\kappa u_*)$ of 1/0.74. The settling velocity is taken from
Soulsby~\cite{soulsby1997}. In this criterion $u_*$ is the peak skin-friction velocity of the waves alone, $u_* = (f_w'/2)^{1/2}U$, where $U$ is the local near-bed orbital velocity amplitude after radial shielding and $f_w'$ is the wave friction factor of Madsen for the grain roughness $2.5D_{50}$ with $D_{50}$ = 0.15~mm. The current is not included. This $D_{50}$ is the reference fine sand of the exceedance fractions and was not measured in a nest. The local stress contrast across one wavelength was taken from the first-order solution of Laplace's equation for the stream function over a sinusoidal bed, $\psi = U[z - a\exp(-kz)\cos kx]$ with $a = \eta/2$ and $k = 2\pi/\lambda$. The near-bed speed is $U(1 + ka)$ over the crest and $U(1 - ka)$ over the trough, with $ka = \pi\eta/\lambda$. The ratio of the two is taken as the ratio of friction velocities, equation~(\ref{eq:contrast}). The solution is applied locally, with the ridge height and wavelength at each radius. It neglects the curvature of the radial array and the convergence of the flow. The disturbance decays upward over a height $\lambda/2\pi$, which over the ridge zone is less than the wake thickness $\delta_w$. The near-bed speeds are linear in $\pi\eta/\lambda$, so the ratio is kept as a geometric index of the contrast. In the Stokes range the settling velocity is proportional to $D^2$, so each cut-off is proportional to $u_*^{1/2}$. The width of the window is then the square root of the ratio in equation~(\ref{eq:contrast}) and does not depend on the level of forcing. With the full settling law, which rises more slowly than $D^2$ for coarser grains, the cut-off rises faster than $u_*^{1/2}$. The Rouse number serves only as a threshold between bedload and suspension. Convective suspension over vortex ripples is treated in Supplementary Note~2.

The coefficients are not fitted to grain size. The window is evaluated with the ninetieth percentile of the near-bed orbital velocity of the waves admitted through the aperture in each depth interval, 0.26--0.47~m~s$^{-1}$, with their median peak period of 9.1~s and the ripple-family roughness with $C$ = 27.7. The admitted waves reach the site during 10.8\% of the breeding season. They are the waves counted in the exceedance fractions above, which are fractions of the whole season. The ninetieth-percentile state is therefore reached for 1.1\% of the season. At the ninetieth-percentile forcing the crest friction velocity, 17.5--40.6~mm~s$^{-1}$ over the ridge zone, exceeds the critical value for the upper edge of the window, 11.7--12.2~mm~s$^{-1}$, at every radius. At the radius of the widest window the trough friction velocity, 7.9--13.5~mm~s$^{-1}$, against a critical value for the local lower edge of 10.6--11.1~mm~s$^{-1}$, stays below it in the two deeper intervals. The equilibrium ripple field used for comparison takes its steepness from the field relation of Nielsen~\cite{nielsen1992}, $\eta/\lambda = 0.342 - 0.34\,\theta_{2.5}^{1/4}$, with $\theta_{2.5}$ the grain-roughness Shields parameter of the same forcing, uniform in radius. The self-similar grader used for comparison is the profile on which the cut-off grain size is a power of radius, $\mathrm{d}\ln D^*/\mathrm{d}\ln r$ constant. This requires the integral in equation~(\ref{eq:transmission}) to grow as $\ln(r_o/r)$, that is $f_w \propto 1/r$. The friction law over this range then gives $k_b \propto r^{-1.92}$ and, with $k_b = C\eta^2/\lambda$ and $\lambda \propto r$, $\eta \propto r^{-0.46}$, scaled to the mean measured ridge height at the mean radius. Any profile $\eta \propto r^{p}$ with $p < 1$ has a steepness decreasing monotonically outward, so the comparison of Fig.~\ref{fig:radial} does not depend on the exponent. The window is evaluated at $\delta_w = h$, inside the calibrated interval of 36--82~mm. Its sensitivity to the Rouse-number convention is given in Supplementary Note~2.

The settling velocity is held constant. Turbulence modifies settling velocity by up to a factor of four in a dimensionless range covering ours, on particles of density ratio 1.4 \cite{baek2025}. The effect of such a modification on the window is derived in Supplementary Note~2.

\subsection{Site geometry and forcing}
The directions open to the Pacific were obtained by casting straight lines from the site at 1$^\circ$ steps, refined to 0.25$^\circ$ across the opening, and intersecting them with the OpenStreetMap coastline vector \cite{osm2026}. The 15 arc-second
global bathymetry \cite{etopo2022} was used only outside the embayment, because inside it places the site in 2~m
of water and spans the 1.15~km entrance with only three to four cells.

Offshore wave climate was taken from two independent reanalyses at the same deep-water
grid point, 21~km from the site on its open side. They are an hourly record for 1995 to 2024
\cite{hersbach2020,era5data,openmeteo} and a three-hourly hindcast for 1990 to 2016
\cite{rascle2013}, which differ in both the
wave model and the atmospheric forcing. The four depth intervals are the depths of the six scanned nests, 13--16 and 21--25~m off Katetsu \cite{kawase2022}, and of the observation areas of the earlier surveys, 18--23~m off Katetsu and 23--28~m off Seisui \cite{kawase2013}. The intervals are kept separate because the near-bed orbital velocity of the waves decreases with depth. The window of Fig.~\ref{fig:window} is computed from the mean measured ridge profile in all four intervals. Transfer from deep water to the nest depths was computed with 48 stationary runs of the spectral wave model SWAN \cite{booij1999}. Waves were admitted to the site only from directions within the aperture. A constant-depth diffraction solution for the 1.15~km aperture, about nine wavelengths wide for a 10~s wave, shows waves from directions outside the aperture to be negligible at the site. Including them raises the exceedance of the transport threshold by at most 0.3 percentage points. Tidal currents were reconstructed from the
four principal constituents of a 42-day moored record in the strait \cite{kunisato2014}.
They were scaled to the site by a ratio of 0.243 read from a published model figure with its
color scale calibrated \cite{hamasoe2019}. The ratio is the least secure input, and the exceedance fractions are given over the bracket 0.22--0.27. The breeding season is taken as April to September, the months of the discoverers' regular surveys \cite{kawase2013}. Nests were observed between 24 March and 20 September in 2011--2013 \cite{kawase2015}. Restricting the wave record to March--September, May--August or June--July changes the combined exceedance by at most 0.4 percentage points.

Near-bed measurements of waves or currents at the site have not been reported. The one current reported at the nesting sites and the forcing in the fortnight before the scans are examined in Supplementary Note~3.

\subsection{Statistics and reproducibility}
The geometry comes from six nests scanned on one day at one site and deposited by Kawase et al.~\cite{kawase2022data}. Each nest is one sample ($n = 6$), and none was excluded. Quantities over the six nests are given as the mean, the standard deviation (SD, with $n - 1$ in the denominator) and the range. Hypothesis tests are not applied. The agreement of the six positions of the steepness maximum is compared with random height profiles (Supplementary Note~1 and Supplementary Fig.~5). Its probability is the fraction of 200,000 sets of six independent profiles per family whose positions agree at least as closely as the measured ones. It depends on the smoothness assumed for the profiles. The center of each nest is located on a random subsample of the scan points. Every random draw uses a fixed seed, so each run of the scripts listed under Code availability gives the same numbers.

\subsection{Use of artificial intelligence}
\begin{figure}[tp]
\centering
\includegraphics[width=\textwidth]{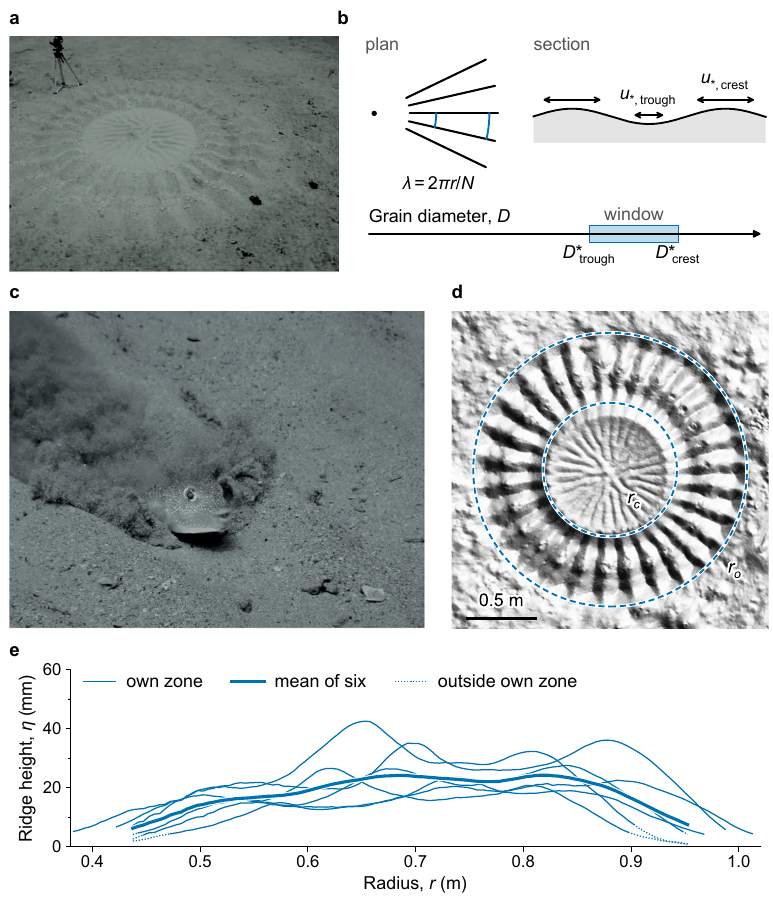}
\caption{\textbf{The structure, the builder, and the measured geometry.}
(a) Completed nest K13, about two meters across, in a frame of the reconstruction video, with a tripod at upper left.
(b) Schematic. In plan, $N$ is fixed, so $\lambda = 2\pi r/N$ shrinks inward. In section, the near-bed flow over crest and trough is drawn with arrow lengths in the ratio $(1+\pi\eta/\lambda)/(1-\pi\eta/\lambda)$ at $\eta/\lambda$ = 0.1, with friction velocities $u_{*,\mathrm{crest}}$ and $u_{*,\mathrm{trough}}$. On the grain-size axis, the confinement window $D^*_\mathrm{trough} < D < D^*_\mathrm{crest}$ is shaded.
(c) A male digging, rolling up fine sand as he swims.
(d) Shaded relief of nest K1 ($N$ = 27) from the deposited model (vertical exaggeration 5); dashed circles mark its own $r_c$ = 0.473~m and $r_o$ = 0.968~m (Table~\ref{tab:morph}); bar, 0.5~m.
(e) Ridge height $\eta(r)$ of the six nests within their own ridge zones (thin; dotted beyond) and their mean (bold).
Panels a and c are Figs.~2b and 1b of Kawase et al.~\cite{kawase2022} (a, video by H.~Kawase), reproduced under CC BY 4.0, cropped and color balanced only; panels d and e derive from their CC0 deposit~\cite{kawase2022data}.}
\label{fig:object}
\end{figure}

\begin{figure}[tp]
\centering
\includegraphics[width=\textwidth]{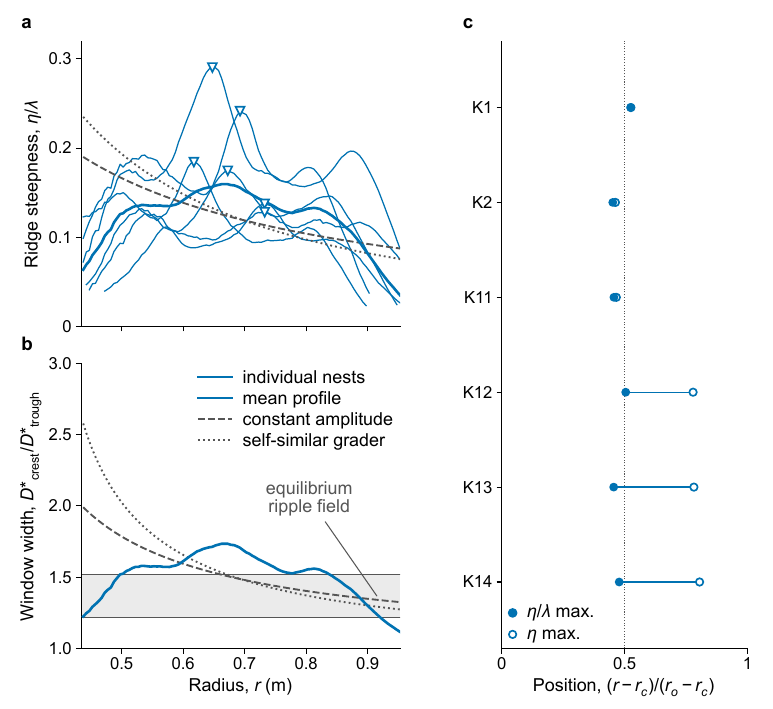}
\caption{\textbf{Radial distribution of ridge steepness and window width in the six scanned nests.}
(a) Local ridge steepness $\eta/\lambda$ against radius $r$. Thin blue lines are the six nests, each drawn over its own ridge zone, and the thick blue line is the mean profile. Open triangles mark the maximum of each nest. The dashed and dotted gray lines are the constant-amplitude and self-similar reference profiles.
(b) Window width $D^*_\mathrm{crest}/D^*_\mathrm{trough}$ in the Stokes range, the square root of equation~(\ref{eq:contrast}), for the mean profile and the two reference profiles. The gray interval, 1.22--1.52, is the window width of an unbounded equilibrium wave-ripple field at the ninetieth-percentile forcing.
(c) Position of the steepness maximum (filled circles) and of the ridge-height maximum (open circles) in each nest, as a fraction of the distance from $r_c$ to $r_o$. The dotted line marks the middle of the zone.}
\label{fig:radial}
\end{figure}

\begin{figure}[tp]
\centering
\includegraphics[width=\textwidth]{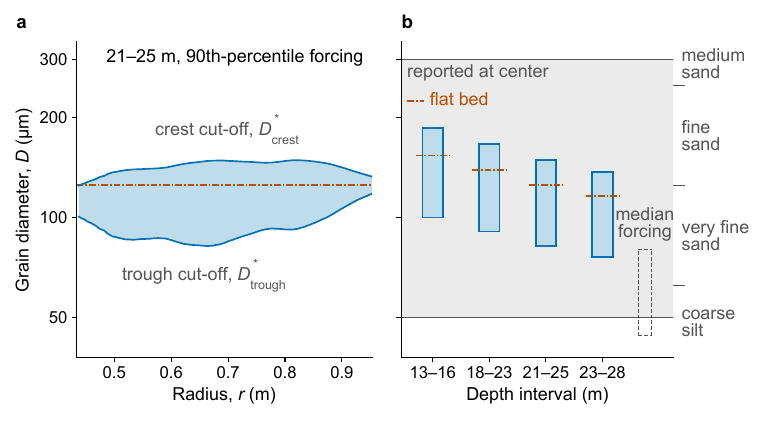}
\caption{\textbf{Predicted confinement window and the grain size reported at the nest center.}
(a) Range of grain diameter between the trough and crest settling limits, $D^*_\mathrm{trough}(r)$ and $D^*_\mathrm{crest}(r)$, across the ridge zone, computed from the mean measured ridge profile with $\delta_w = h$ and $C$ = 27.7 for the 21--25~m interval at the ninetieth-percentile near-bed orbital velocity of the waves admitted through the aperture in the breeding season (blue shading with solid edges). The dash-dot line marks the suspension cut-off over a flat bed under the same forcing. (b) Extent of the window over the ridge zone in each depth interval (bars), the flat-bed cut-off in each interval (dash-dot ticks) and, at 23--28~m, the window at median forcing (dashed open bar). The gray shading spans the 50--300~$\mu$m range reported for the sand accumulating at the nest center \cite{kawase2016kaken}, which has no radial position. The scale at right gives the Wentworth classes, with ticks at 62.5, 125 and 250~$\mu$m.}
\label{fig:window}
\end{figure}

The conception of this work (the question, the physical model and the structure of the argument) is the author's. AI tools (Anthropic Claude; Fable~5.2, Opus~5.5 and earlier Claude models) were used, under the author's direction, for drafting of the analysis and figure scripts, derivation checking, numerical verification, and drafting of English prose; all results and text were reviewed and are the responsibility of the author. Generative-AI images were not used; every figure was drawn by those scripts, and the photographs in Fig.~\ref{fig:object} are reproduced from ref.~\cite{kawase2022}. The scripts include tests against limiting cases, the bedform test was calibrated with synthetic ripples of known height, the extracted relief was compared with the published relief of the same nests, the numbers in the text were recomputed at full precision and rounded once, and the key numbers were recomputed independently.

\backmatter

\bmhead{Data availability}
The three-dimensional nest models analyzed here are the structure-from-motion scans
deposited under CC0 by Kawase et al.~\cite{kawase2022data}. The offshore wave climate is taken from
the ERA5 reanalysis \cite{era5data}, retrieved through the Open-Meteo Marine API
\cite{openmeteo}, and from the IOWAGA WAVEWATCH~III hindcast \cite{rascle2013}, retrieved
from the Ifremer THREDDS server. Tidal constituents and the site attenuation ratio are
taken from Kunisato et al.~\cite{kunisato2014} and Hamasoe et al.~\cite{hamasoe2019}, and the Australian substrate from
Geoscience Australia~\cite{ausseabed}. The site geometry uses
the OpenStreetMap coastline \cite{osm2026}, Esri World Imagery \cite{esri2026}, and the
ETOPO 2022 relief model \cite{etopo2022}, and the spectral wave runs use SWAN version 41.51
\cite{booij1999}. The per-nest morphometry and the extracted radial profiles are archived at Zenodo \cite{park2026data}.

\bmhead{Code availability}
The analysis and figure scripts behind every number, table and figure in this paper are archived at Zenodo \cite{park2026data}, with a list mapping each script to the sections and figures it supports.

\bmhead{Acknowledgements}
The author acknowledges
support from the Institute of Engineering Research and the Institute for Peace and
Unification Studies at Seoul National University.

\bmhead{Funding}
This research was supported by the National Research Foundation of Korea (NRF) grant
funded by the Korea government (MSIT) (No. RS-2026-25475182).
The funder had no role in the design of the study, the analysis, the decision to publish or the preparation of the manuscript.

\bmhead{Author contributions}
Y.S.P. conceived the study, performed the analysis and wrote the manuscript.

\bmhead{Competing interests}
The author declares no competing interests.


\clearpage
\clearpage
\clearpage

\end{document}